\documentstyle[12pt]{article}

\begin{document}


\def\a{\alpha}
\def\b{\beta}
\def\c{\varepsilon}
\def\d{\delta}
\def\e{\epsilon}
\def\f{\phi}
\def\g{\gamma}
\def\h{\theta}
\def\k{\kappa}
\def\l{\lambda}
\def\m{\mu}
\def\n{\nu}
\def\p{\psi}
\def\q{\partial}
\def\r{\rho}
\def\s{\sigma}
\def\t{\tau}
\def\u{\upsilon}
\def\v{\varphi}
\def\w{\omega}
\def\x{\xi}
\def\y{\eta}
\def\z{\zeta}
\def\D{\Delta}
\def\G{\Gamma}
\def\L{\Lambda}
\def\F{\Phi}
\def\P{\Psi}
\def\S{\Sigma}

\def\o{\over}

\def\IJMP{Int.~J.~Mod.~Phys. }
\def\MPL{Mod.~Phys.~Lett. }
\def\NP{Nucl.~Phys. }
\def\PL{Phys.~Lett. }
\def\PR{Phys.~Rev. }
\def\PRL{Phys.~Rev.~Lett. }
\def\PTP{Prog.~Theor.~Phys. }
\def\ZP{Z.~Phys. }

\def\beq{\begin{equation}}
\def\eeq{\end{equation}}


\title{
  \begin{flushright}
    \large UT-824
  \end{flushright}
  \vspace{5ex}
  $R$-Invariant Dilaton Fixing}
\author{Izawa K.-I.~and T.~Yanagida \\
  \\  Department of Physics, University of Tokyo, \\
  Tokyo 113-0033, Japan}
\date{September, 1998}
\maketitle

\begin{abstract}
We consider dilaton stabilization with $R$ invariance,
which insures a vanishing cosmological constant
at the scale of stabilization.
We construct a few models which accommodate weak gauge couplings
with large or small gauge groups.
Matter condensation plays a central role in the dilaton stabilization.
\end{abstract}

\newpage

\section{Introduction}

It is expected that a unified theory of all interactions (yet to be
discovered) will allow description based on effective field theory
at low energy.
One possible feature of a low-energy effective theory is
that its coupling constants should be determined dynamically,
since a unified theory may well have no arbitrary parameters
to provide {\it ad hoc} coupling constants.   
In fact, superstring theory as a candidate of such a unified theory yields,
at least perturbatively, this type of effective field
theory, which generally contains, among others, the dilaton field 
as a modulus to determine the overall coupling strength.

In this paper, we consider an $R$-invariant stabilization
of the vacuum expectation value of the dilaton.%
\footnote{This dilaton is not necessarily
the dilaton in perturbative string theory
\cite{Kap}.
We only require that the vacuum expectation value of this dilaton
determines the gauge couplings at the unification scale.}
An advantage of the $R$ invariance is that the cosmological constant
at the stabilization scale is naturally vanishing in the stabilized vacuum,%
\footnote{The $R$ symmetry for model building in relation with
a vanishing cosmological constant is considered in Ref.\cite{Yan}.
The vanishing of the cosmological constant with supersymmetry breaking
is not achieved without an appropriate $R$ breaking \cite{Nel}.}
in contrast to schemes without $R$ invariance
\cite{Kra,Ban}.
We provide a few models of dilaton fixing
with large or small gauge groups,
where the dilaton is stabilized in the vacuum corresponding to
weak gauge couplings.

\section{Dilaton Stabilizer}

Let us consider a superpotential of the form
\beq
 W = Xf(\F),
 \label{SP}
\eeq
where $X$ denotes a chiral superfield carrying $R$ charge 2
and $f(\F)$ is a function of
the dilaton supermultiplet $\F$ with vanishing $R$ charge.
This superpotential can be generated by dynamics of gauge interactions.

The potential in supergravity is given by
\begin{equation}
 V = e^{K} (K_{AB} F^A F^{B*} - 3|W|^2),
 \label{EPOT}
\end{equation}
where $K$ is a K{\" a}hler potential,
$K_{AB}$ denotes the inverse of the matrix
\beq
 {\q^2 K \o \q \f_A \q \f_B^*},
\eeq
with $\f_A = X, \F$,
and $F^A$ is given by
\beq
 F^A = {\q W \o \q \f_A} + {\q K \o \q \f_A}W.
\eeq
Here we set the gravitational scale equal to one
(in the Einstein frame).

Judicious choice of the function $f(\F)$ can stabilize the dilaton
$\F$ in the supersymmetric $R$-invariant vacuum:
\beq
 \langle X \rangle = 0, \quad f(\langle \F \rangle) = 0,
 \quad \langle V \rangle = 0,
 \label{VAC}
\eeq
where $\langle F^A \rangle = \langle W \rangle = 0$.%
\footnote{This is a local minimum of the potential
with unbroken supersymmetry
\cite{Wei}.
Whether or not this is the global minimum of the potential
depends on the K{\" a}hler potential.}

In the following sections, we discuss dynamics giving rise to
suitable superpotentials for the dilaton $\F$.

\section{Dynamical Preliminaries}

We first discuss the dynamics causing matter condensation,
since this is crucial
for dilaton stabilization in our models. We adopt an Sp$(N)$
gauge theory with $2(N+1)$ chiral superfields $Q_i$
in the fundamental $2N$-dimensional representation,%
\footnote{Similar dynamics were considered in Ref.\cite{Lou}
with the dilaton field. The dynamics without the dilaton were
investigated in Ref.\cite{Sei} and utilized to construct
vector-like models of dynamical supersymmetry breaking \cite{Int}.}
where $i$ is a flavor index ($i=1,\cdots,2(N+1)$)
and the gauge index is omitted.
Without a superpotential, this theory has a flavor SU$(2(N+1))_F$ symmetry.
This SU$(2(N+1))_F$ symmetry is explicitly broken down to a flavor Sp$(N+1)_F$
by a superpotential in our models.
That is, we add gauge singlets $Z^a$ ($a=1,\cdots,N(2N+3)$)
to obtain the tree-level superpotential
\cite{Hot}
\footnote{This superpotential may provide an inflationary sector
in our models
\cite{Iza}.}
\begin{eqnarray}
 W_0 = \lambda Z^a (QQ)_a,
\end{eqnarray}
where $(QQ)_a$ denotes
a flavor $N(2N+3)$-plet
of Sp$(N+1)_F$ given by a suitable combination of gauge invariants $Q_iQ_j$.

The effective superpotential
which may describe the dynamics of the Sp$(N)$ gauge interaction
\cite{Lou,Sei}
is given by
\begin{eqnarray}
 W_{eff}=S({\rm Pf} V_{ij} - \Lambda^{2(N+1)}) + \lambda Z^a V_a
\label{dynamical_potential}
\end{eqnarray}
in terms of low-energy degrees of freedom
\begin{eqnarray}
 V_{ij} \sim Q_iQ_j, \quad V_a \sim (QQ)_a,
\end{eqnarray}
where $S$ is an additional chiral superfield and
$\L$ denotes a dynamical scale of the gauge interaction,
which is given by $\L = \exp(-8\pi^2\F/b)$ with $b=2(N+1)$.

The effective superpotential Eq.(\ref{dynamical_potential}) implies that,
among the gauge invariants $Q_iQ_j$, the Sp$(N+1)_F$ singlet $(QQ)$
condenses as
%
$\langle (QQ) \rangle = \Lambda^2$
%
with supersymmetry unbroken in the corresponding vacuum
\cite{Hot}.
Since the vacuum preserves the flavor Sp$(N+1)_F$ symmetry, we have no 
massless Nambu-Goldstone multiplets. The absence of flat directions
except for the dilaton $\F$ at this stage
is crucial for achieving dilaton stabilization
in the subsequent sections.

\section{Simple Models}

Let us take a single ${\rm Sp}(N)$ gauge theory
with $2(N + 1)$ chiral superfields $Q$ in the fundamental
representation, as above. We omit the flavor index of $Q$ here and henceforth.
The superpotential of our simple model is given by
\beq
 W = \l Z^{a}(QQ)_{a}
     + X(f_1(QQ) - f_2(QQ)^2).
 \label{SP1}
\eeq
For simplicity, we have neglected higher-order terms in $(QQ)$,
whose inclusion does not change our conclusion qualitatively.

Integrating out the Sp$(N)$ gauge interaction along with the gauge singlets
$Z^a$,
we obtain a condensation $\langle (QQ) \rangle = \L^2$,
as shown in the previous section,
where $\L = \exp(-8\pi^2\F/b)$ and $b=2(N+1)$.
Then the effective superpotential of the model is given by
Eq.(\ref{SP}) with the function
\beq
 f(\F) = f_1 e^{-{16\pi^2 \o b}\F} - f_2 e^{-{32\pi^2 \o b}\F}.
\eeq

The $R$-invariant vacuum is given by
\beq
 \langle X \rangle = 0, \quad
 e^{-{16\pi^2 \o b}\langle \F \rangle} = {f_1 \o f_2}.
\eeq
To obtain a desired value ${\rm Re} \langle \F \rangle \simeq 2$
of the dilaton condensation,%
\footnote{The gauge coupling constant $g$ is given by
$g^2 = 1/{\rm Re} \langle \F \rangle$. Hence
${\rm Re} \langle \F \rangle \simeq 2$ corresponds to the gauge coupling
constants of the standard model gauge groups determined at the grand
unification scale of order $10^{16}$ GeV.}
we need a large $N$ $(\simeq 150)$ for $|f_1/f_2| \simeq e^{-1}$.  
We note that only one gauge group is sufficient to fix the dilaton $\F$.

\section{Semisimple Models}

We adopt two ${\rm Sp}(N_i)$ gauge theories
with $2(N_i + 1)$ chiral superfields $Q_i$ in the fundamental representations
($i = 1, 2$).
We consider two types of semisimple models.

$1)$
The superpotential of the first model is given by
\beq
 W = \l_1 Z_1^{a_1}(Q_1Q_1)_{a_1} + \l_2 Z_2^{a_2}(Q_2Q_2)_{a_2}
     + X(f_1(Q_1Q_1) - f_2(Q_2Q_2)).
 \label{SP2}
\eeq

Integrating out the ${\rm Sp}(N_1) \times {\rm Sp}(N_2)$ gauge interactions
along with the gauge singlets $Z_i^{a_i}$,
we obtain condensations $\langle (Q_iQ_i) \rangle = \L_i^2$ for $i = 1, 2$.
Here $\L_i$ denotes the dynamical scales of the ${\rm Sp}(N_i)$
gauge interactions,
which are given by $\L_i = \exp(-8\pi^2\F/b_i)$ and $b_i = 2(N_i+1)$.
Then the effective superpotential of the model is given by
Eq.(\ref{SP}) with the function
\beq
 f(\F) = f_1 e^{-{16\pi^2 \o b_1}\F} - f_2 e^{-{16\pi^2 \o b_2}\F}.
\eeq

The $R$-invariant vacuum is given by
\beq
 \langle X \rangle = 0, \quad
 e^{-{16\pi^2 \o b}\langle \F \rangle} = {f_1 \o f_2},
\eeq
where $b = b_1b_2/(b_1-b_2)$.
The desired value ${\rm Re} \langle \F \rangle \simeq 2$
for $|f_1/f_2| \simeq e^{-1}$
is realized by $N_1 = 12$ and $N_2 = 11$, for example.

$2)$
The superpotential of the second model is given by
\beq
 W = \sum_{i=1,2} \l_i Z_i^{a_1}(Q_iQ_i)_{a_i}
     + Y(f'_1 \P^{n_1} - f_1 (Q_1Q_1))
     + X(f'_2 \P^{n_2} - f_2 (Q_2Q_2)),
 \label{SP3}
\eeq
where $Y$ and $\P$ are singlet chiral superfields.

Integrating out the ${\rm Sp}(N_1) \times {\rm Sp}(N_2)$ gauge interactions
along with the gauge singlets $Z_i^{a_i}$,
$Y$ and $\P$,
we obtain the effective superpotential of the model
in the form Eq.(\ref{SP}) with the function
\beq
 f(\F) = f'_2 \left({f_1 \o f'_1}\right)^{n_2 \o n_1}
         e^{-{16\pi^2 n_2 \o n_1b_1}\F} - f_2 e^{-{16\pi^2 \o b_2}\F}.
\eeq

The $R$-invariant vacuum is given by
\beq
 \langle X \rangle = 0, \quad
 e^{-{16\pi^2 \o b}\langle \F \rangle}
 = {f_2 \o f'_2}\left({f'_1 \o f_1}\right)^{n_2 \o n_1},
\eeq
where $b = n_1b_1b_2/(n_2b_2-n_1b_1)$.
The desired value ${\rm Re} \langle \F \rangle \simeq 2$
for $|f'_1/f_1| \simeq |f_2/f'_2| \simeq e^{-1}$
is realized by $N_1 = 1$, $N_2 = 2$, $n_1 = 13$ and
$n_2 = 9$, for example.
The sizes of the gauge groups can be minimal in the present model,
though small gauge groups yield a small dilaton mass,
which might be undesirable phenomenologically.

\section{Conclusion}

We have introduced a dilaton stabilizer $X$
with the superpotential Eq.(\ref{SP}),
which gives the supersymmetric $R$-invariant vacuum Eq.(\ref{VAC})
with a vanishing cosmological constant.%
\footnote{The stabilizer $X$ may cause a hybrid inflation in conjunction with
the dilaton $\F$.}
We have utilized matter condensation rather than gaugino
condensation to stabilize the dilaton $\F$. This is a crucial difference
from the previous works
\cite{Kra,Ban}.
We have considered a few dynamical models Eqs.(\ref{SP1}), (\ref{SP2})
and (\ref{SP3}),
where the dilaton is stabilized at the vacuum expectation value
corresponding to weak gauge couplings.

Once the dilaton is fixed,
various dynamical models without the dilaton
can be used. For example, dynamical supersymmetry
breaking (see Ref.\cite{Int})
and dynamical inflation (see Ref.\cite{Iza}) may be realized
with the fixed dilaton.
The scale of dynamical inflation is determined by that of its gauge
dynamics. The runaway vacuum ${\rm Re} \F \rightarrow \infty$ induces no
inflation, while the stabilized vacuum can incorporate dynamical
inflation if the stabilization scale is sufficiently large.
In this case, the universe may be naturally dominated
by the stabilized
vacuum through the inflationary dynamics under chaotic initial conditions.



\begin{thebibliography}{99}

\bibitem{Kap}
  V.~Kaplunovsky and J.~Louis, \PL {\bf B417} (1998) 45.

\bibitem{Yan}
  Izawa~K.-I. and T.~Yanagida, \PL {\bf B393} (1997) 331;
  \PTP {\bf 97} (1997) 913; \PTP {\bf 99} (1998) 423; \\
  T.~Yanagida, \PL {\bf B400} (1997) 109.

\bibitem{Nel}
  T.~Banks, D.B.~Kaplan and A.E.~Nelson, \PR {\bf D49} (1994) 779; \\
  J.~Bagger, E.~Poppitz and L.~Randall, \NP {\bf B426} (1994) 3.

\bibitem{Kra}
  N.V.~Krasnikov, \PL {\bf B193} (1987) 37; \\
  J.A.~Casas, Z.~Lalak, C.~Mu{\~ n}oz and G.G.~Ross, \NP {\bf B347}
  (1990) 243; \\
  T.R.~Taylor, \PL {\bf B252} (1990) 59; \\
  B.~de~Carlos, J.A.~Casas and C.~Mu{\~ n}oz, \NP {\bf B399} (1993) 623.

\bibitem{Ban}
  T.~Banks and M.~Dine, \PR {\bf D50} (1994) 7454; \\
  P.~Bin{\' e}truy, M.K.~Gaillard and Y.-Y.~Wu, \NP {\bf B481} (1996)
  109; {\bf B493} (1997) 27; \PL {\bf B412} (1997) 288; \\
  J.A.~Casas, \PL {\bf B384} (1996) 103; \\
  T.~Barreiro, B.~de~Carlos and E.J.~Copeland, \PR {\bf 57} (1998) 7354.

\bibitem{Wei}
  S.~Weinberg, \PRL {\bf 48} (1982) 1776.

\bibitem{Lou}
  V.~Kaplunovsky and J.~Louis, \NP {\bf B422} (1994) 57; \\
  E.~Dudas, \PL {\bf B375} (1996) 189; \\
  G.~Dvali and Z.~Kakushadze, \PL {\bf B417} (1998) 50; \\
  M.~Klein and J.~Louis, hep-th/9803143.

\bibitem{Sei}
  N.~Seiberg, \PR {\bf D49} (1994) 6857; \\
  K.~Intriligator and P.~Pouliot, \PL {\bf B353} (1995) 471.

\bibitem{Int}
  Izawa~K.-I. and T.~Yanagida, \PTP {\bf 95} (1996) 829; \\
  K.~Intriligator and S.~Thomas, \NP {\bf B473} (1996) 121. 

\bibitem{Hot}
  T.~Hotta, Izawa~K.-I.~and T.~Yanagida, \PR {\bf D55} (1997) 415; \\
  Izawa~K.-I., Y.~Nomura, K.~Tobe and T.~Yanagida,
  \PR {\bf D56} (1997) 2886.

\bibitem{Iza}
  Izawa~K.-I., \PTP {\bf 99} (1998) 157.

\end{thebibliography}
\end{document}